\documentclass[aps,prb,floatfix,twocolumn,showpacs]{revtex4}
\usepackage{epsfig}
\usepackage{amsmath}
\usepackage{amssymb}
\usepackage{amsfonts}
\usepackage{ dsfont }

\usepackage[usenames,dvipsnames]{color}

\begin{document}

\hsize\textwidth\columnwidth\hsize\csname@twocolumnfalse\endcsname

\bibliographystyle{plain}

\title{NMR Response of  Nuclear Spin Helix in Quantum Wires \\ with Hyperfine and Spin-Orbit Interaction}

\author{Peter Stano$^{1,2}$ and Daniel Loss$^{1,3}$}
\affiliation{ $^1$RIKEN Center for Emergent Matter Science, 2-1 Hirosawa, Wako, Saitama, 351-0198 Japan}
\affiliation{$^2$Institute of Physics, Slovak Academy of Sciences, 845 11 Bratislava, Slovakia}
\affiliation{$^3$Department of Physics, Klingerbergstrasse 82, University of Basel, Switzerland}

\vskip1.5truecm
\begin{abstract}
We calculate the nuclear magnetic resonance (NMR) response of a quantum wire where at low temperature a self-sustained electron-nuclear spin order is created. Our model includes the electron mediated Ruderman-Kittel-Kasuya-Yosida (RKKY) exchange, electron spin-orbit interactions, nuclear dipolar interactions, and the static and oscillating NMR fields, all of which play an essential role. The paramagnet to helimagnet transition in the nuclear system is reflected in an unusual response: it absorbs at a frequency given by the internal RKKY exchange field, rather than the external static field, whereas the latter leads to a splitting of the resonance peak. 
\end{abstract}
\maketitle

\section{Introduction}
Semiconductor nanowires offer unusual flexibility, which makes them attractive platforms for  spintronics,\cite{zutic2004:RMP, fabian2007:APS} and quantum information.\cite{kloeffel2013:ARCMP}  Materials with sizable spin-orbit interaction (SOI), like n-InAs or p-SiGe nanowires, seem promising for spin qubits in gated quantum dots,\cite{trif2008:PRB,nadj-perge2010:N, petersson2012:N,kloeffel2013:PRB}
while nuclear-spin-free materials such as $^{29}$Si or $^{13}$C
are pursued\cite{zwanenburg2009:NL}  to overcome the coherence limit of GaAs,\cite{petta2005:S} alternative to  decoupling schemes,\cite{bluhm2010:N} dynamical nuclear polarization,\cite{bracker2005:PRL} or state engineering.\cite{reilly2008:S, rudner2007:PRL} Apart from material considerations, designs based on tunable one-dimensional conductors are suitable for realization of topologically nontrivial excitations such as Majorana fermions,\cite{lutchyn2010:PRL,oreg2010:PRL,mourik2012:S,deng2012:NL,das2012:NP,rokhinson2012:NP,alicea2012:RPP} non-Abelian fractionally 
charged fermions,\cite{klinovaja2012:PRL} and parafermions.\cite{oreg2014:PRB,klinovaja2014a:PRL,klinovaja2014:PRB}

Strong interaction effects are ubiquitous for one-dimensional electrons, boosting the response to charge and spin excitations.\cite{stano2013:PRB} In the electron-mediated Ruderman-Kittel-Kasuya-Yosida (RKKY) interaction\cite{ruderman1954:PR,kasuya1956:PTP,yosida1957:PR} 
between nuclear spins, this is reflected as an enhanced  peak at finite momentum, which was predicted to drive nuclear spins into a helical ferromagnet, at a critical temperature 
of about 100 mK in GaAs at strong interactions.\cite{braunecker2009:PRB}
Similar behavior was recently predicted to occur in coupled-wire models.\cite{meng2014:EPJB}
Establishing such nuclear order would suppress spin noise,\cite{schliemann2003:JPCM} stabilize fractional fermions,\cite{klinovaja2012:PRL} and provide a tune-free platform for Majorana fermions.\cite{klinovaja2013a:PRL,braunecker2013:PRL,vazifeh2013:PRL}

The helical order gives rise to a nuclear Overhauser field which acts back on the electrons  and opens a gap in one of the originally spin-degenerate  subbands. Importantly, such a gap reduces the conductance
from 2 to 1 $e/h^2$ upon lowering the temperature.\cite{braunecker2009:PRB} This behavior was recently observed  in GaAs quantum wires by Scheller {\it et al.},\cite{scheller2014:PRL} who were able to rule out all other
known mechanisms except nuclear spins. However, the evidence for the latter is still indirect.
This drawback has motivated us to search for additional characteristics of the helical nuclear order
that will allow one to confirm or  exclude 
nuclear spins as possible explanation for the  anomalous conductance behavior observed in Ref.~\onlinecite{scheller2014:PRL}.

We find that nuclear magnetic resonance (NMR) can offer a direct experimental test.\cite{zhukov2014:PRB}
Indeed, as we show below, the NMR response of a helically ordered nuclear state of a one-dimensional conductor is completely different from the one of a para- or  ferromagnetic state. Though we use a GaAs semiconducting quantum wire as a specific example, our results apply to a broad range of electron phases, both ungapped and gapped.\cite{klinovaja2013a:PRL}
The only assumption which our results require, beyond the one that the helical nuclear order is established by mediating electrons, is that the SOI wave vector is larger than the width of the peak
of the RKKY interaction in momentum space.
It is well satisfied in our GaAs example. 
Because the  driving NMR field couples to excitations with a $\it finite$ wave vector in a helimagnet, and because of the narrowly peaked RKKY exchange, 
even a small shift of the wave vector, as induced by the SOI, makes the corresponding NMR frequency be set by the internal RKKY field, rather than by the external static $B$ field, 
in stark contrast to the usual case of para- and ferromagnets.

The outline of the paper is as follows. We first introduce a model of the coupled electron-nuclear system (Sec.~II), comprising the RKKY and dipolar interactions between nuclei, the spin-orbit interaction of electrons, and the Zeeman terms of the NMR fields. We then find the ground state of the model Hamiltonian in Sec.~III and its low-energy excitations (magnons) in Sec.~IV. From there we derive the NMR absorption characteristics (the resonant frequency and dispersion) in Sec.~V.

\section{Model}

We consider a quasi-one-dimensional semiconductor, such as a GaAs wire, or a carbon nanotube, in which electrons (labeled by $n$) are governed by the Hamiltonian
\begin{equation}
H_E = \sum_n H_0(x_n, p_n) + \sum_{n,m} U(x_n-x_m) + \sum_{n,i} A {\bf S}_{ni} \cdot {\bf I}_i,
\label{eq:electrons}
\end{equation}
comprising the electron single-particle part $H_0=p^2/2m + H_{so} $, 
interactions $U$, and coupling  to nuclear spins with hyperfine coupling constant $A$. Here,
 $p$ is the momentum and $m$ the effective mass of the electron,
$H_{so}= (\hbar p / 2m l_{\rm so}) {\bf n} \cdot \boldsymbol{\sigma}$ is the SOI, defined by the SOI axis (a unit vector ${\bf n}$) and length $l_{\rm so}$, and the Pauli matrix $ \boldsymbol{\sigma}$ (times $\hbar/2$) denotes the electron spin.
The nuclear spin operator ${\bf I}_i$, $i=1,\ldots,N,$ stands for a  collective spin of the cross-sectional transverse plane of the wire, a sum of $N_\perp = \pi R^2 a \rho$
aligned nuclear spins, each of magnitude $I_{3/2}$, with $R$ the wire radius, $a$ the lattice constant, and $\rho$ the nuclear-spin density. This one-dimensional model of effective spins of length $I=N_\perp I_{3/2}$ is discussed in detail elsewhere.\cite{braunecker2009:PRB, meng2014:EPJB} Finally, ${\bf S}_{ni}=\boldsymbol{\sigma}_n\delta(x_n-x_i)/2 \pi R^2 \rho$ is the electron spin density
operator at the $i$th 
transverse plane at $x_i=ia$, and $L=Na$ is the wire length.

We first neglect the SOI. The hyperfine term in Eq.~\eqref{eq:electrons} results in an electron-mediated RKKY interaction,
\begin{equation}
H_{R} = \sum_{i,j} J(x_i-x_j) {\bf I}_i \cdot {\bf I}_j .
\label{eq:RKKY}
\end{equation}
We define the Fourier transform of the RKKY exchange by
\begin{equation}
J_q = \sum_{j=1}^N e^{- i q x_j}J(x_j).
\label{eq:Fourier}
\end{equation}
In one dimension, $J_q=J_{-q}$ develops a pronounced dip
at twice the Fermi wave vector, $|q|=2k_F$, which is strongly renormalized by interactions $U$,
and scales as 
\begin{equation}
J_{2k_F} \propto -\frac{(A/N_\perp)^2}{\hbar v_F/a} \left( \frac{\hbar v_F /a}{k_B T} \right)^{2-2g},
\label{eq:J2kf}
\end{equation}
see Eq.~(29) in Ref.~\onlinecite{braunecker2009:PRB} for the prefactor and other details. The exchange is inversely proportional to the Fermi velocity $v_F=\hbar k_F/m$ and is negative, supporting a ferromagnetic type of order. The electron-electron interactions, characterized by a Luttinger liquid parameter $K_\rho$, reduce the value of the exponent according to $g^2=2K_\rho^2/(1+K_\rho^2)$, which substantially boosts the RKKY interaction upon lowering the temperature $T$ (cf. Fig.~\ref{fig:signal}). 
In a GaAs wire the ordering temperature reaches $\sim 100$ mK at $K_\rho=0.2$.\cite{meng2014:EPJB}

With $J_q$ minimal at a finite wave vector, the ground state of $H_R$ is a helical ferromagnet, with spins oriented according to
${\bf I}_i = \mathcal{R}_{ {\bf h}, 2k_F x_i} [ \boldsymbol{\mathcal{I}} ]$.
Here, $\mathcal{R}_{{\bf h},\alpha}$ is the operator of a three-dimensional rotation by an angle $\alpha$ around a unit vector ${\bf h}$. The spins rotate in the helical plane (perpendicular to ${\bf h}$) as one moves along the wire, with period $\pi/k_F$.
At this point, the problem has full spin rotation symmetry, which means that the directions of  ${\bf h}$ and the reference spin $\boldsymbol{\mathcal{I}}$ may be chosen arbitrarily, as long as $\boldsymbol{\mathcal{I}} \perp {\bf h}$. This symmetry will be broken in a realistic finite wire with anisotropies. Instead, here we consider additional interactions: the nuclear dipolar coupling and the electron SOI. Though much weaker than the RKKY interaction [so that the dipolar effects on the NMR absorption characteristics (the resonant frequency and dispersion) are negligible], together they pin the helical plane.

Summing the nuclear-spin dipole-dipole energies pairwise, we derive, 
by averaging the interaction over the wire cross section,
the dipolar interaction Hamiltonian for the effective spins to be
\begin{equation}
H_{dd}= \sum_{i,j}  d(x_i-x_j)  {\bf I}_i \cdot D  {\bf I}_j \equiv \sum_{i,j}  {\bf I}_i \cdot D_{ij}  {\bf I}_j\, ,
\label{eq:dipole}
\end{equation}
where
\begin{equation}
d(x)=\mu_0 \mu^2 N_\perp \rho \frac{R^2a}{2[x^2+R^2]^{3/2}}\, ,
\label{eq:dx}
\end{equation}
with $\mu_0$ being the vacuum permeability and $\mu$ the average nuclear magnetic moment. The form of the $3\times 3$ matrix $D={\rm diag} ( -1, 1/2, 1/2)$ follows from the dipole-dipole interaction symmetry and the assumed wire cylindrical symmetry. Since $d(x)$ is exact in the limits $R\gg |x_i-x_j|$ and $R\ll |x_i-x_j|$, we expect $H_{dd}$ to be an excellent approximation. The matrix $D$ is anisotropic---it is energetically favored for the spins to point along the wire axis. Breaking the symmetry, it restricts the helical plane to contain the wire axis.

Consider now the SOI term $H_{so}$. In higher dimensions, its effect is involved and may induce a helical order itself.\cite{maleyev2006:PRB,kirkpatrick2008:PRB,loss2011:PRL}
Here, we account for it by a unitary transformation $H_E \to U H_E U^\dagger$, with  $U=\exp[(i x / 2l_{so} ) {\bf n}\cdot \boldsymbol{\sigma}]$, which removes the SOI from Eq.~\eqref{eq:electrons} except for the replacement 
\begin{equation}
{\bf I}_i \to {\bf I}^{so}_i = \mathcal{R}_{{\bf n},-x_i/l_{so}} [{\bf I}_i],
\label{eq:SOC frame}
\end{equation}
where we used the identity $U \boldsymbol{\sigma} \cdot {\bf I} \, U^\dagger = \boldsymbol{\sigma} \cdot \mathcal{R}^{T} [{\bf I}]$, with $\mathcal{R}$  performing the same rotation on vectors as $U$ does on spinors. The exchange interaction Eq.~\eqref{eq:RKKY} is then unchanged by the SOI, if written in the rotated reference frame, that is, with the replacement Eq.~\eqref{eq:SOC frame}. The SOI field thus defines another direction, ${\bf n}$, to which the helical order axis is fixed uniquely, as we will see.

The Zeeman energy of the NMR static (${\bf B}$) and driving [${\bf b}(t)$] magnetic fields completes our model,
\begin{equation}
H_s+H_d = -\sum_i \mu {\bf B} \cdot {\bf I}_i - \sum_i \mu {\bf b}(t) \cdot {\bf I}_i.
\label{eq:NMR fields}
\end{equation}
We neglect their coupling to the electron subsystem as we will consider small static field and off-resonant (to the electron Zeeman energy) driving field. Not to complicate the analysis, we consider a Faraday configuration, ${\bf b}\perp{\bf B}$, usual for NMR. 

Even though our results are applicable to various systems, we stick to GaAs as a typical example.\cite{meng2014:EPJB} Here, $A=90\,\mu$eV, $I_{Ga}=I_{As}=I_{3/2}=3/2$,  $a=5.65$\AA, $\rho=8/a^3$, $\mu\approx 1.2\mu_N$ with $\mu_N$ the nuclear magneton. We further use $k_F=2\times10^5$ m/s, $R=5$ nm, $L\sim 2-40\,\mu$m, 
$K_\rho=0.2$, and $T=30$ mK as representative values for the results given below.

To investigate the NMR response, we first find the ground state of the effective spins, in the presence of the exchange, dipolar, and SOI, and the static NMR field. It happens to be a conical spiral. We then use the Holstein-Primakoff ansatz to calculate the excitations (magnons) spectrum. Finally, we use the linear response formalism to calculate the moments of the absorption line. 

\section{Ground state}

Without dipolar effects,  the ground state energy per effective nuclear spin follows from Eqs.~\eqref{eq:RKKY}-\eqref{eq:J2kf},
\begin{equation}
E_{R} = J_{2k_F} N_\perp^2 I_{3/2}^2 \approx - 46\,{\rm meV}.
\end{equation}
We now consider dipolar interactions. In the SOI rotated  frame, Eq.~\eqref{eq:SOC frame}, the matrix $D_{ij}$ in Eq.~\eqref{eq:dipole} becomes 
\begin{equation}
D^{so}_{ij}=\mathcal{R}_{{\bf n},-x_i/l_{so}}  D_{ij}   \mathcal{R}_{{\bf n},x_j/l_{so}.}
\label{eq:rotated dipole}
\end{equation}
Since $R \ll l_{so}$ and the dipolar interaction falls off fast for $|x_i-x_j|\gg R$, we can set $x_i=x_j$ in Eq.~\eqref{eq:rotated dipole}. For Dresselhaus SOI with ${\bf n}||x$, the rotations in Eq.~\eqref{eq:rotated dipole} commute with the matrix $D_{ij}$ and thus cancel, and there is no pinning of the helical plane by SOI. On the other hand, 
for wires grown along [110],  the Dresselhaus SOI ${\bf n} || y$, as is the case for a Rashba SOI 
induced by the field of the heterostructure interface (z axis).
The SOI-induced rotation of the reference frame is then nontrivial.  To proceed, we average Eq.~\eqref{eq:rotated dipole} over the wire coordinate. This is a useful technical simplification, as it restores the translational invariance,  without affecting results substantially, as it only slightly changes the small dipolar contribution to the magnon energies. The resulting averaged matrix $\overline{D_{ij}^{so}} \propto {\rm diag} (-1/4,1/2,-1/4)$, leads to an energy gain in the plane perpendicular to ${\bf n}$, which is therefore the helical plane. Transforming back to the crystal coordinate system, we further see that it is beneficial if the SOI induced rotation slows down, rather than speeds up, the helical rotation. 
The complete undo of the helical rotation at the commensurability condition, 
$2k_F^{so} \equiv 2k_F-1/l_{so}=0$, can result in fractional and Majorana fermions.\cite{klinovaja2012:PRL}
We conclude that the SOI and dipolar couplings together fix ${\bf h}$ to be antiparallel to ${\bf n}$. The dipolar energy per effective spin follows as
\begin{equation}
E_{dd} = -I_{3/2}^2 d_{2k_F^{so}}/4 \approx -7 \, {\rm neV}, 
\end{equation}
with $d_q$ the Fourier transform, defined as in Eq.~\eqref{eq:Fourier}, of $d(x)$ from Eq.~\eqref{eq:dx}. The dipolar energy is orders of magnitude smaller than the RKKY energy. 

Consider now a weak static NMR field that is applied perpendicular to the helical plane. The helically ordered spins will be deflected and lie on a cone,\cite{michael2010:PRB} which amounts to rotating the reference spin $\boldsymbol{\mathcal{I}}$ towards the static field $\boldsymbol{\mathcal{I}} (B) = \mathcal{R}_{z, \theta} [\boldsymbol{\mathcal{I}}(B=0)]$. To find the rotation angle $\theta$, we minimize the ground state energy $\langle H_{R}+H_s \rangle$ with respect to $\theta$ and get
\begin{equation}
\sin \theta = \frac{\mu B}{2N_\perp I_{3/2}(J_0-J_{2k_F})},
\label{eq:theta}
\end{equation}
a deflection proportional to the static field $B$. The Zeeman energy per effective nuclear spin is quadratic in $B$,
\begin{equation}
E_z = - \mu B I \sin\theta = -\frac{(\mu B)^2}{2(J_0-J_{2k_F})} \approx - 93\,{\rm neV}\, ,
\label{eq:Zeeman}
\end{equation}
where we used $B=1$ T to get the numerical value. The conical ansatz is in fact dictated by the problem symmetry, and is valid at strong static fields as well. Increasing $B$, the helical magnet smoothly deforms into a uniform ferromagnet, the transition completed at a critical field defined by $\sin\theta=\pi/2$, Eq.~\eqref{eq:theta} giving $B_{\rm c} \sim 10^3$ T.

Had we assumed $\bf B$ within the helical plane, the ground state properties are qualitatively the same (unlike in higher dimensions\cite{kishine2009:PRB}): the small field deflection is linear and the energy gain quadratic in $B$. The differences are mainly practical; in this case again the translational symmetry is lost which greatly complicates calculations. Therefore, we set ${\bf B} || y$, expecting the results for other directions to be qualitatively the same. As a minimum, this option is possible to choose in experiments, as the axis $y$ is not arbitrary, but is perpendicular to the Rashba electric field and the wire axis. 

\section{Magnon spectrum}

Summarizing the above, the following Hamiltonian collects essential ingredients,
\begin{equation}
H=\sum_{i,j} J_{ij} {\bf I}^{so}_i \cdot {\bf I}^{so}_j+ \sum_{i,j}  {\bf I}^{so}_{i} \cdot \overline{D_{ij}^{so}}  {\bf I}^{so}_j- \sum_i  \mu {\bf B} \cdot {\bf I}_i.
\label{eq:final Hamiltonian}
\end{equation}
No less important is the ansatz for the ground state,
\begin{equation}
{\bf I}^{so}_i=\mathcal{R}_{{\bf h},2k_F x_i}  \mathcal{R}_{z,\theta} [\boldsymbol{\mathcal{I}}_i],
\end{equation}
with ${\bf n} || y || {\bf B}$, $\theta$ given by Eq.~\eqref{eq:theta}, and the ground state direction chosen arbitrarily within the helical plane, e.g. $\boldsymbol{\mathcal{I}}_i=(I,0,0)$. To diagonalize $H$ we use the Holstein-Primakoff ansatz, $\mathcal{I}_q^x=IN\delta_{q,0}-\sum_p a^\dagger_{p-q} a_p$, $\mathcal{I}_q^y=\sqrt{NI/2}(a^\dagger_{-q}+a_q)$, $\mathcal{I}_q^z= i \sqrt{NI/2}(a^\dagger_{-q}-a_q)$ for the Fourier transform of $\boldsymbol{\mathcal{I}}_i$ defined as in Eq.~\eqref{eq:Fourier}. The introduced boson operators have standard commutation relations $[a_p, a^\dagger_q]=\delta_{p,q}$. The ansatz introduces an error $O(1/I)$, which is small since $I=N_\perp I_{3/2}\gg 1$. 

After lengthy but straightforward calculation we get in the harmonic approximation (constant omitted)
\begin{equation}
H\approx \sum_q (\hbar \omega_q +\epsilon_q^- I \sin\theta) b_q^\dagger b_q,
\label{eq:diagonal Hamiltonian}
\end{equation}
in terms of the new boson operators, a result of a Bogoliubov transformation $b_q=u_q a_q - v_q q^\dagger_{-q}$ with $u_q^2-v_q^2=1$, and
\begin{equation}
\epsilon_q^\pm =J_{2k_F+q}\pm J_{2k_F-q} -( d_{2k_F^{so}+q} \pm  d_{2k_F^{so}-q})/4.
\end{equation}
The system excitations (magnons) have definite momentum and energies with a direction insensitive part,
\begin{equation}
(\hbar \omega_q)^2 = I^2 (\epsilon^+_q-\epsilon^+_0) [ \epsilon^+_q-\epsilon^+_0+(2J_q+d_q-\epsilon^+_q)\cos^2\theta],
\label{eq:energy symmetric}
\end{equation}
and a directional splitting $\epsilon_q^- I \sin\theta$, induced by the external $B$ field [{\it cf.}~Eq.~(\ref{eq:Zeeman})] if the exchange (dipolar) interactions are asymmetric with respect to the $q=2k_F$ ($2k_F^{so}$) point.

Zero-momentum magnons have zero energy, since $\omega_0=0=\epsilon_0^-$. As these represent a rotation of the whole system, the zero gap indicates a rotational symmetry. Indeed, there is one, as all characteristic vectors are collinear, ${\bf n}||{\bf B}||{\bf h}$. We expect additional anisotropies, like spin impurities or wire asymmetry, to select a specific direction for $\boldsymbol{\mathcal{I}}$ in the helical plane, equivalent to a small energy for $q=0$ excitations. Expanding in $q$, a linear energy dispersion follows from Eqs.~\eqref{eq:diagonal Hamiltonian}-\eqref{eq:energy symmetric} for long wavelength magnons. In an extended system, this would mean the order can not exist at any finite temperature, in accordance with general theorems.\cite{mermin1966:PRL, loss2011:PRL} The linearization, however, is valid only for momenta smaller than the width of the dip in $J$ around its minimum. This width is actually smaller than the inverse of the wire length, so that the lowest allowed $q$ magnon is already gapped appreciably and the order can be established. (See Ref.~\onlinecite{meng2014:EPJB} for a detailed critical temperature analysis.)

\section{Absorption line moments}

\begin{figure}
\includegraphics[width=0.45\textwidth]{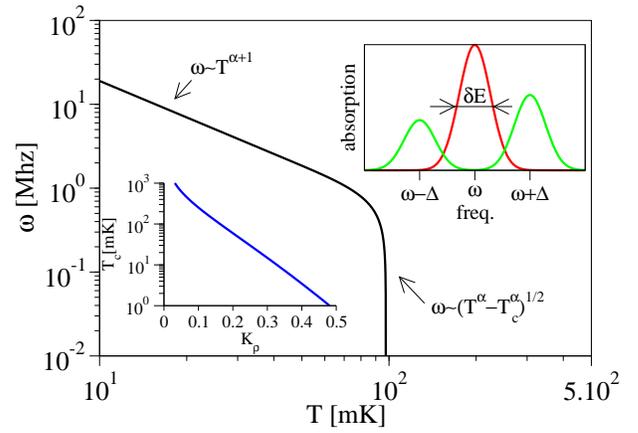}
\caption{(Color online) Absorption signal characteristics. Main (black): Center of the absorption line as a function of temperature $T$. Plot of the the power-law dependence of the NMR resonance Eq.~(\ref{eq:main}) in the regions far away (left) and close (right) to the critical temperature $T_c$, with $\alpha=2g-3$. 
Right inset: Absorption as a function of NMR frequency (line profile), for zero (single peak, red; with $\hbar \omega\approx -2IJ_{2k_F}$) and finite (double peak, green; with $\Delta=|\epsilon^-_{2k_F^{so}}I \sin \theta|$) external $B$ field. Left inset (blue): 
$T_c$ as function of the Luttinger liquid  parameter $0<K_\rho \leq 1$ ($\equiv$ noninteracting limit). 
  \label{fig:signal} }
\end{figure}

Finally, we turn to the NMR absorption. Splitting the driving field Hamiltonian into  energy increasing and decreasing parts, $H_d=H_d^++H_d^-$, 
with $H_d^+=(H_d^-)^\dagger$, 
we find (using $v_{2k_F^{so}}\ll 1$)
\begin{equation}
H_d^- \approx i \mu b(t) \sqrt{NI/8} [b_{2k_F^{so}}(1+\sin\theta)+b_{-2k_F^{so}}(1-\sin\theta)].
\label{eq:drive}
\end{equation}
The spatially uniform driving field excites magnons with $q=\pm 2k_F^{so}$, which gives rise to two resonant energies,
\begin{equation}
E_\pm \approx -2IJ_{2k_F} \sqrt{1-(J_{2k_F^{so}}/J_{2k_F})\cos^2\theta} \pm \epsilon_{2k_F^{so}}^- I \sin\theta,
\label{eq:main}
\end{equation}
where we assumed $|J_{2k_F}|$ is much larger than other energies, including $|J_{2k_F^{so}}|$. The above constitutes our main result and shows how the helical order manifests itself in an NMR experiment. 
Most notably, the NMR resonance Eq.~(\ref{eq:main})  is centered at the internal field $J_{2k_F}$, rather than at the external one. The latter leads to a small redshift ($\propto \cos^2\theta$) and, more importantly, to a peak splitting ($\propto \sin\theta$). In addition, Eq.~\eqref{eq:drive} shows that the two peaks differ in height, proportional to $(1\pm\sin\theta)^2$; see Fig.~\ref{fig:signal}. Whether this splitting is observable depends on the linewidth, which arises from the so far neglected magnon-magnon interactions.

By going beyond the harmonic approximation Eq.~\eqref{eq:diagonal Hamiltonian}, we take these terms now into account, and calculate the linewidth $\delta E=\surd\langle E^2\rangle-\langle E \rangle^2$ 
using the linear response formula for the absorption line second moment ($\rho_{0}$ is the canonical density matrix)\cite{slichter} 
\begin{equation}
\langle E^2 \rangle = {\rm Tr} (\rho_{0} [ [H,H_d^-],[H_d^+,H] ] ) / {\rm Tr} (\rho_{0} [H_d^-,H_d^+] ).
\end{equation}
We find the leading contribution to be 
\begin{equation}
\delta E = \frac{\cos^2\theta}{2N\hbar^2}  \sum_q (|\epsilon^-_q|^2 +  |2J_q+d_q-\epsilon^+_q|^2 \sin^2\theta),
\label{eq:delta E}
\end{equation}
for zero magnon occupations. 
The evaluation of  $\delta E$ is not possible in closed form, as the dependence of $J$ on $q$ (away from the peak) is not known well. 
However, we note that the absorption, being proportional to  $N^0$ (after replacing the sum by an integral), is not extensive and, therefore, is negligible. This is no longer so in the case of macroscopic magnon populations, where further terms contribute. As their  evaluation is   again possible only with the knowledge of the form of $J_q$ for all momenta, we only note that for our regime of primary interest, namely for temperatures smaller than the critical temperature,
these contributions are exponentially small. We conclude that, with the ingredients included in our model, the NMR absorption splitting in the external field is observable. Whether this is really so will depend on additional contributions to the linewidth, such as those from the electron system, or the nuclear and impurity spin environment. Our predictions are summarized in Fig.~\ref{fig:signal}.

We note that for the result given in Eq.~\eqref{eq:main}, the SOI is crucial. Namely, for  negligible SOI, $k_F^{so}\to k_F$, Eq.~\eqref{eq:main} gives $E_\pm = \mu B$ [valid for large enough $B$; the exact form as well as the small-$B$ limit follows from Eq.~\eqref{eq:energy symmetric}]. The crossover SOI length between these two limits is given by the inverse width of the peak in $J$ around $q=2k_F$, expected to be much larger than micrometers, the scale for the SOI length in GaAs.\cite{zumbuhl2014:CM} Concluding, it is the SOI which makes the NMR response of the helical order different than a paramagnetic or a ferromagnetic one.   

Finally,  in a  recent  landmark experiment,\cite{peddibhotla2014:NP} NMR responses on nanostructures including nanowires have been  detected successfully via cantilever techniques. This has  opened up an entire new class of nanoscale systems for NMR-based experiments, and makes us believe that the NMR effects predicted here are well within experimental reach.

\acknowledgments
We acknoweldge support by APVV-0808-12 (QIMABOS) (P.S.), SCIEX, Swiss NF, and NCCR QSIT.

\bibliographystyle{apsrev}
\bibliography{../../../Xfiles/docs/articles/references/quantum_dot}

\end{document}